\documentclass[11pt]{article}
\usepackage{amsmath,amsfonts,epsfig} 
\usepackage{amssymb}

\numberwithin{equation}{section}
\title{Fast Simulation 
\\
of Multicomponent Dynamic Systems}
\author{Boris D. Lubachevsky\\
\\
{\em bdl@bell-labs.com}\\
\\
Bell Laboratories\\
600 Mountain Avenue\\
Murray Hill, New Jersey\\
}
\date{}
\begin{document}
\maketitle
\begin{abstract}
A computer simulation
has to be fast to be helpful, if it
is employed to study the behavior of
a multicomponent dynamic system.
This paper discusses modeling concepts 
and algorithmic techniques
useful for creating such fast simulations.
Concrete examples of simulations
that range from econometric modeling to communications
to material science are used to illustrate
these techniques and concepts.
The algorithmic and modeling methods discussed 
include event-driven processing, 
``anticipating'' data structures, 
and ``lazy'' evaluation, 
Poisson dispenser,
parallel processing by 
cautious advancements
and by
synchronous relaxations.
The paper gives examples of how
these techniques and models are employed in assessing
efficiency of capacity management methods
in wireless and wired networks, in studies of 
magnetization, crystalline structure, and sediment formation
in material science, in studies of competition in economics.

\end{abstract}
\section{Introduction}\label{s:intro}
Dynamic systems with many interacting components
are encountered in areas ranging from econometric modeling
to communications to material science.
Individual components and their local interactions
may or may not be complex,
but their mere multiplicity often creates a complexity
which needs to be addressed.

As an example, consider the Flexchan feature,
introduced in new releases of the TDMA wireless system.
According to the Flexchan method,
each base station that serves
a cell in the service area,
periodically checks interference level on different channels
and
dynamically rearranges the
channels in the order of the 
increase of the expected interference.
A new service request (a new call
or a handoff request from a mobile 
with the call in progress)
is allocated according to an accepted strategy
taking into account availability of free channels and the 
segregation order.
For instance, a ``greedy''
strategy places the call on the unoccupied channel 
with the least expected interference.

All the channel capacity of the system is potentially available
to each cell in Flexchan
in contrast with Fixed allocation schemes,
that are largely in use today.
The latter prespecifies a partition of capacity among the cells
in anticipation of the traffic.
Under the Flexchan, cells themselves negotiate
the capacity,
dynamically forming an allocation pattern
in response to the actual traffic.
Thus, the Flexchan eliminates the manual planning,
reduces the operating cost and
presumably increases capacity and QoS.

Does it?
Will the channel segregation adapt
to the traffic,
or maybe it will instead 
oscillate somehow, with base stations ``fighting''
each other for the capacity?
If it does adapt,
how long would the adaptation process take
depending on algorithm parameters
such as the frequency of checking the
interference by base stations?
Short of actual system deployment,
only simulations can answer these questions.

To be convincing, the simulation
should be dynamic with base stations asynchronously
working out their Flexchan routine while users
are moving in the service area.
Many base stations
should be represented
in order to demonstrate the algorithm viability;
it is possible to imagine how the algorithm would work
for just a few base stations,
but break down 
for, say, a hundred base stations.
It becomes obvious 
that a crucial element of this simulation
is the computing efficiency of the simulation algorithm.

This paper reviews algorithmic and programming techniques
which were used to answer 
the posed questions by simulation\cite{BGKKLS97}.
The program was written so that,
say, for about 100 base stations and 1000 active mobiles,
simulating one hour of operation
requires only several hours of processing 
on a single PC or workstation.

Similar algorithmic concepts and
techniques for improving efficiency 
of {\em discrete event} simulations
recur in diverse applications such as:

- an econometric model of telephone companies, like AT\&T and MCI, 
fighting among themselves 
for the residential telephone market quotas.
To attract the customers,
the players introduce various payment plans,
like ``Friends and Family'' or ``Volume discount.''
A question may be: which policy/discount works
better given the same cost for the player.
The customers behave randomly, their response is staggered,
but they tend to behave in accordance with their
individual economic interests.
It was possible to simulate such systems with more than 100,000
customers with individualized connections and ``habits''
in calling each other
during several simulated years 
with a single run taking a few minutes on a PC.

- a dynamic routing in a wired network, like
the long distance AT\&T network, where the use of 
an ``anticipating'' data structure
allows one to cut processing time of simulating one operating
hour from a hundred hours to under a ten hours
and where further speed improvement using parallel
processing shrinks the processing time to a few minutes.

- multiparticle studies in computational physics,
and material science, such as a model of magnetization,
a model of particle deposition, a study of impurity-perturbed
crystals. Here ``lazy'' evaluation, Poisson dispensing,
and parallel processing lead to several order of magnitude
speed improvements. 
Some 
simulations
previously thought impossible,
as they would take
years to complete under old techniques, 
with the 
new programming techniques
move to the category of possible ones,
those that can be completed in several hours.

This paper is organized as follows.
Sections \ref{s:edr}, 
\ref{s:lazy}, and
\ref{s:poisson} present techniques and concepts
for simulating multicomponent systems on a uniprocessor. 
The advantages and difficulties of event-driven processing are discussed in
Section \ref{s:edr},
the balance between ``lazy'' and anticipatory methods of evaluation 
is explained in Section \ref{s:lazy},
and a general method of event-driven simulation which avoids, and which is
faster than the event list method is introduced in Section \ref{s:poisson}.
Sections \ref{s:sequent} and
\ref{s:relax} present technique for simulating efficiently
large multicomponent systems on a multiprocessor.
A ``cautious'' technique which avoids speculative computations
is discussed in Section \ref{s:sequent}.
If the latter technique is not feasible,
one may resort to 
the synchronous relaxation method, presented in Section \ref{s:relax}.
The lessons learned in the
course of the practical application of the discussed methods
are discussed in Conclusion.
\section{Time-driven vs. event-driven simulations}\label{s:edr}

A time-driven description of a dynamic system involves the global clock.
The time-driven computer model keeps in memory the global state
of the system and modifies the entire state 
as the global time advances in finite increments.
Although time-driven models are intuitive
and convenient to think of,
one should try to avoid them in simulations
because of their computational inefficiency.
Converting the model from the time-driven
to an event-driven form, 
whenever such conversion is possible,
is a single best improvement to a computer simulation.
That is a classical textbook recommendation.
In a general case,
it is not known, though, how 
to do
the conversion.
\subsection{Time-driven one-dimensional billiards}
As an example of the conversion, 
consider simulation of a collection
of chaotically colliding billiard balls.
Despite its toyish appearance,
the ``billiards'' simulation technically is non-trivial
and has serious applications.

\begin{figure}[h]
\centerline{\psfig{file=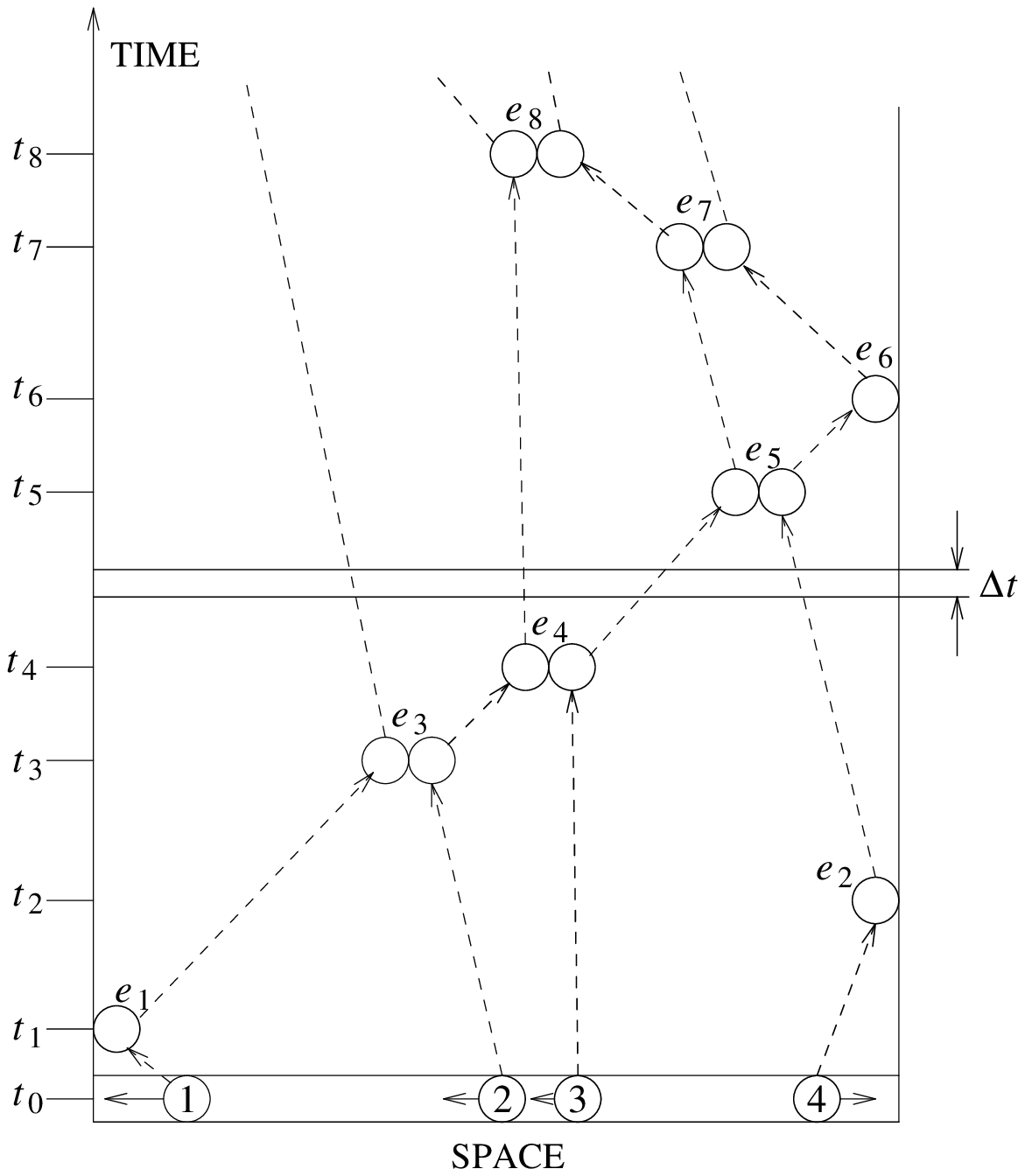,width=6.1in}}
\caption{Billiards in one dimension}
\label{gutter}
\end{figure}

For simplicity consider billiards in one dimension.
We may imagine a gutter bounded from both ends,
which contains $N$ absolutely hard 
elastic balls of equal mass and size.
The width of the gutter is just enough to let the balls move
without friction in one dimension.
The gutter is placed horizontally
which excludes the effect of gravitation on the ball movements.
\marginpar{\em Figure \ref{gutter}}
The gutter filled with
$N=4$ balls 
is shown at the bottom 
of Figure \ref{gutter} which also
presents
the initial time-space trajectories of the balls for times $t > 0$.
The trajectories are indicated by dashed lines,
they are initiated at $t = 0$
by the shown positions and velocities.

To simulate the system evolution in a time-driven way,
we integrate equations of motion along
small time segments of duration $\Delta t$.
Thus, starting with time $0$
we advance the state to time
$\Delta t$, 
then to $2 \Delta t$,
then to $3 \Delta t$,
and so on.

If a ball experiences no collision during 
the advancement interval \mbox{$(t,t+\Delta t)$},
then, because there is no friction, we have
\begin{equation}
\label{ballstep}
x(t+ \Delta t) = x(t) + v(t)  \Delta t ,\ \ \ \  v(t+ \Delta t) = v(t)
\end{equation}
where $x(t)$ is the one-dimensional position coordinate
and $v(t)$ is the velocity of the ball (center) at time $t$.

At each step the time-driven method monitors 
distances between components that might come in contact.
Ball-ball collisions are detected by monitoring 
distances $d(i,i+1)$ between centers of balls $i$ and $i+1$.
Such a collision occurs during
time interval $(t , t + \Delta t )$
if $d(i,i+1)$ becomes smaller that ball diameter $D$ at time $t + \Delta$,
while $d(i,i+1) \ge D$ at time $t$.
Once detected, the collision is processed by 
exchanging velocities of the balls at time $t + \Delta t$:
\begin{equation}
\label{collision}
{v_i }^{new} = {v_{i+1} }^{old} \ , \ \ \ \ \ \ 
{v_{i+1} }^{new} = {v_i }^{old}
\end{equation}

It follows, that for a single ball the algorithm detects and processes
at most one collision during a $\Delta t$ step. 
With a large $\Delta t$, the algorithm may fail
to represent interdependent collisions
that follow over a single $\Delta t$ step.
Hence, $\Delta t$ determines the accuracy of the simulation;
smaller $\Delta t$ higher the accuracy.
\subsection{An application: impurity perturbed crystal}
A high accuracy is required in many applications
of the billiards simulation,
for instance, for
generating a pattern of impurity perturbed
crystal.
A material scientist might be interested to know
the geometry of displacements of particles in
a crystal formed of identical particles 
with a tiny fraction of inserted isolated 
particles of a different kind, 
the impurity.

\begin{figure}[h]
\centerline{\psfig{file=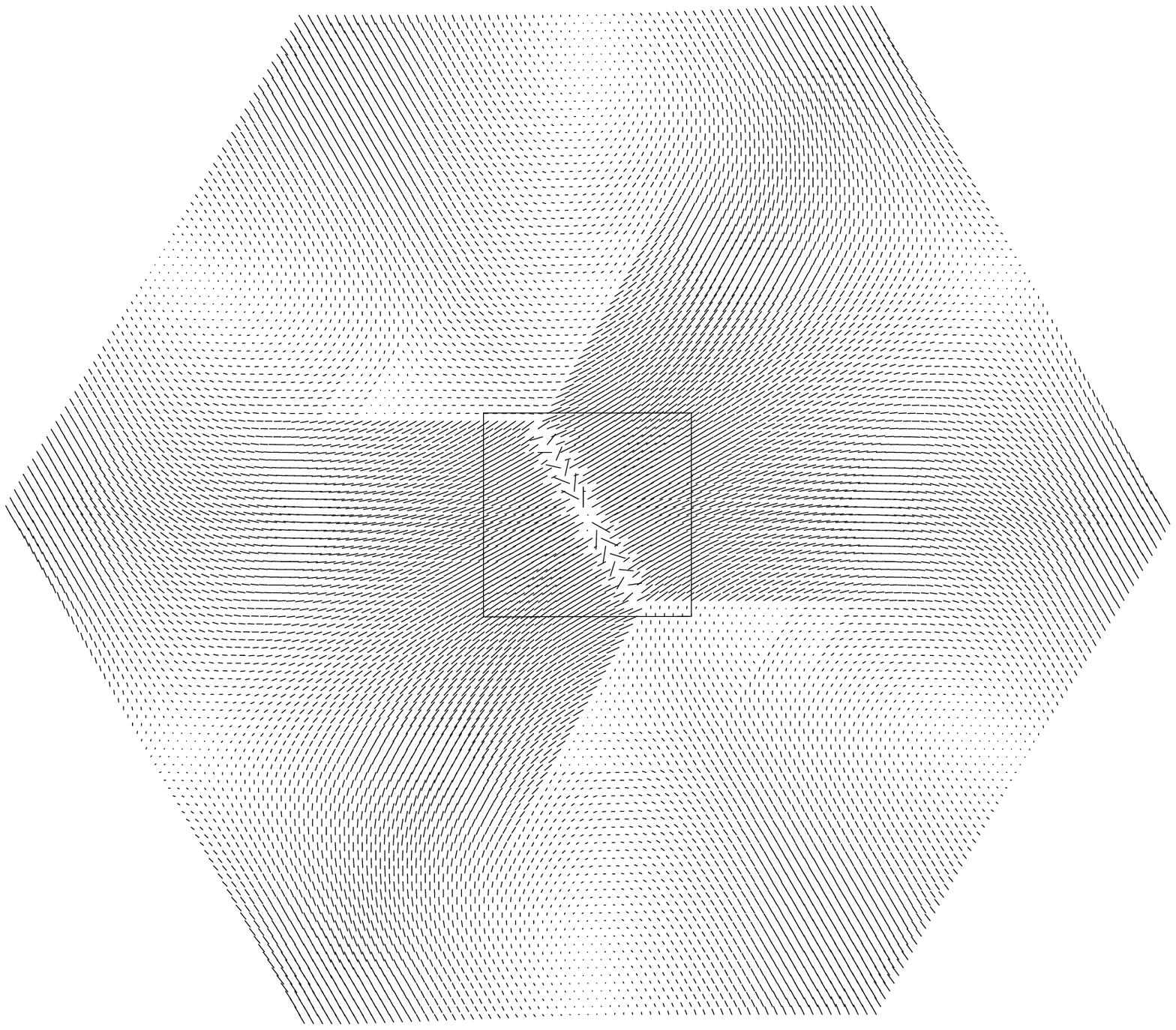,width=6.1in}}
\caption{Displacement pattern of particles in a hexagonal crystal
perturbed by a larger impurity particle in the center.
The assembly consists of about 11000 particles.
The outlined central square is reproduced in Figure \ref{mosaic} }
\label{hex}
\end{figure}

\begin{figure}[h]
\centerline{\psfig{file=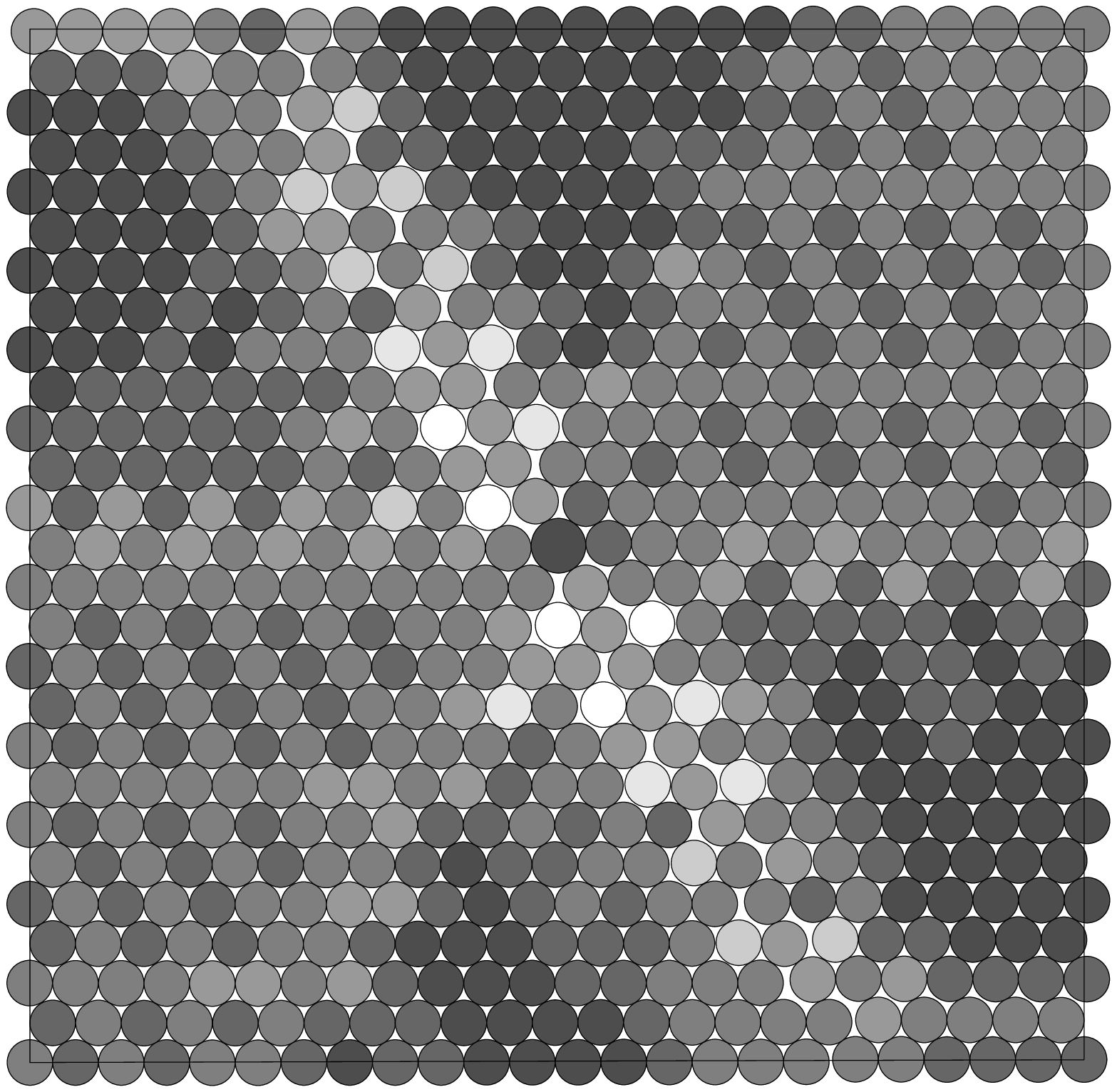,width=6.1in}}
\caption{Particle arrangement in the central square oulined
in Figure \ref{hex}.
The impurity particle in the center is 20\% larger that the rest.
Particles have been classified by number of contacts with neighbors
according to shading:
those that have more contacts with neighbors are darker,
white particles are ``rattlers'' with no contacts }
\label{mosaic}
\end{figure}

The mechanism of generating a displacement pattern in two
dimensions, see Figures \ref{hex} and \ref{mosaic},
can be modeled by billiards as follow.
\marginpar{\em Figures \ref{hex},\ref{mosaic}}
We introduce an impurity {\em larger} particle 
into a perfect hexagonal crystal
of circular particles of common diameter
packed inside a bounded geometric shape,
say, inside a hexagon.
To fit the larger diameter circle we loosen the assembly 
by simultaneously decreasing the diameters of all the circles.
The decrease of the particles matches
the excess size of the impurity so that we then able to
substitute a single regular particle
with
the larger impurity particle without creating particle overlaps.
The impurity particle is in
the center in Figures \ref{hex} and \ref{mosaic}.
We then randomly assign a velocity to each particle
and let them all uniformly increase in diameter, so that
the ratios of the diameters of the 
particles remain constant.
``Swelling'' particles chaotically move in different directions
colliding with each other 
thereby negotiating the limited free space
inside the bounded region
until the final ``jammed'' state delivers
the sought pattern.

Figure \ref{hex} represents 
vectors of 
displacement
of the particles from their respective original positions
in the unperturbed crystal
and Figure \ref{mosaic} represents the particles themselves
in the central square fragment 
of the pattern in Figure \ref{hex}.
The pattern is robust.
It repeats itself under different
initial conditions within a range of parameters such as
a ratio of the larger particle
to the regular ones and a rate of particle
expansion.

At the final phase of the expansion,
when the particles are almost ``jammed''
a quick sequence of interdependent collisions has a high chance
to occur within the array of 11,000 particle.
The accuracy required
for capturing this interdependence is high
and the $\Delta t$ required for it is small.
Should the $\Delta t$ be not small enough, the simulation
fails to produce the correct pattern.

Unfortunately, 
the accuracy does not agree with the efficiency of computations.
In a typical $\Delta t$ step, 
if the $\Delta t$ is small, 
the state of most particles is advanced according to \eqref{ballstep}
with no collision.
Such is the case
for the $\Delta t$ step shown in Figure \ref{gutter}.
The no-collision advancements waste computing power:
generating the pattern similar to the
one shown in Figures \ref{hex} and \ref{mosaic} would take
more than a year on a personal computer.

That is an estimate though, because
the time-driven method 
was never tried successfully in this example.
Instead, the pattern was generated using an event-driven
method and the computations for a pattern like the one in
Figures \ref{hex} and \ref{mosaic} were typically completed
in under 10 hours of processing on a PC.
\subsection{Event-driven billiards}
The event-driven method assumes a discrete event model
of the system, wherein the system and its components
change their states instantaneously at discrete times;
those changes are called the {\em events}.
The state remains constant on the intervals between the events.
The trajectory of the system is a directed acyclic
{\em event dependency graph}. 
The nodes of it are the events,
and the links represent cause-effect relations in pairs of events.

This graph can be also viewed as a data-flow diagram.
To each node-event corresponds the event {\em descriptor},
which consists of the time of the event and
of the specification of the state change represented by the event.
If events 
$e_1 , e_2 , .... ,e_k$ 
are all the immediate causes
of event $e$, then the descriptor of event $e$
is a function of the descriptors of
$e_1 , e_2 , .... ,e_k$.
Generating the descriptor of event $e$
from the descriptors of $e$'s causes
is {\em processing} event $e$.

Figure \ref{gutter} 
represents not only the 4-ball system trajectory
but can be also viewed as the event dependency
graph corresponding to this trajectory.
The events $e_i$, $i = 1,2,...$,
are ball-ball and ball-wall collisions.
The arrows from an event to an event along a ball trajectory
can be seen as cause-effect links on the  event dependency graph
or as data-dependency links of the data-flow diagram.
An event descriptor here consists of:
the collision time, the positions and velocities
of the involved balls immediately after the collision.

One can check, for instance, that (the descriptor of) event $e_5$,
a collision of balls 3 and 4, is a function of
(the descriptors of) event $e_4$, a collision of 2 and 3,
and $e_2$, a bounce of 4 against the right end of the gutter.
Namely, the time component of $e_5$ descriptor,
$t(e_5) = t_5$, is computed as
\begin{equation}
\label{event5t}
t_5 = t_4 + \frac {x_4 (t_2) - x_3 (t_4) + v_4 (t_2)\  (t_4 - t_2) - D} {v_4 (t_2) - v_3 (t_4) }
\end{equation}
where $x_i (t)$ and $v_i (t)$ 
are position and velocity of ball $i$, $i = 1,2,3,4$,
at time $t$,
$t_i = t(e_i )$ is the time of event $e_i$ as indicated on the
time axis in Figure \ref{gutter},
and $D$ is the ball diameter. 
As claimed, the right hand-side
in \eqref{event5t} is a function of
event $e_2$ descriptor, including $t_2$,
$x_4 (t_2)$, and  (negative) $v_4 (t_2)$,
and event $e_4$ descriptor, including $t_4$,
$x_3 (t_4)$, and (positive) $v_3 (t_4)$.

Producing the simulated history of a system under study
is the goal of a simulation. The goal is achieved by
processing events along the history.
But how do we know which events to process?
Neither the event descriptors nor the topology of
the event dependency graph is known in advance.
The event-driven simulation must do both
construct the event dependency graph and process the events on it.
The former activity, is also referred to as event {\em scheduling}.

The task of designing an efficient scheduling 
usually constitutes the main difficulty of recasting a time-driven
model into an event-driven form.
A well known mechanism of scheduling an event-driven simulation
on a uniprocessor is
a {\em queue of events}.
With the queue one is able to insert future events in the schedule,
and to retrieve the next event for processing.
A popular data structure \cite{K68} for the event queue is
{\em heap} with which either
operation is performed in order of $\text{log} N$ computing operations
where $N$ is the number of events in the queue.
By contrast, a straightforward scheduling method would need 
order of $N$ operations either for retrieving or for inserting or for both.
If $N = 11,000$, say, then slow down may be substantial.

It should be stressed though, that handling event queue efficiently
does not cover the entire task of efficient event-driven simulation
\cite{PAS}.
The simulationist also has
to impose appropriate modeling assumptions.
For example, in the simulation of billiards,
were the balls not absolutely hard, 
the collisions would not be instantaneous.
Instead of a simple velocity exchange in \eqref{collision},
this would force a (slow) time-driven evaluation of the collisions.
Equally important for the speed of computation
is the 
assumption of
the motion without friction.
With friction, the ball motions would not be 
integrable as simply as in \eqref{ballstep}.
\subsection{Mobility of users in wireless simulations}
The wireless simulations discussed in Section \ref{s:intro}
are naturally of a discrete event type,
events being call arrival, termination, handoff, interference
measurement, adjustment of the channel segregation order.
Among few exceptions is the users' mobility.
The users may move along curvy trajectories
with variable speeds. It is decided, however,
that the simulated users move 
with constant velocities along 
straight line segments.
As in the billiards model,
this assumption
leads to an event-driven processing.
Validity of the simulation results is not
seriously affected 
because a curvy motion with variable velocity can be
approximated with a sequence of the motions of the considered type.

Selecting a model agreeable with
the event-driven mechanism and using
an efficient event queue may still be not enough
for efficiency of an event-driven simulation.
The following sections will discuss
more techniques and approaches toward the goal of a fast 
multicomponent simulation.

\section{Anticipatory vs. ``lazy'' evaluation}\label{s:lazy}

A simulation 
may generate queries concerning the system state.
The anticipatory method can be defined
as constructing answers
for questions which have not yet been asked.
It can be contrasted with a ``lazy'' 
evaluation method according to which generation of the answer
is delayed until
the latest possible moment when the query
is issued.
The anticipatory evaluation tends to be event-driven,
while the lazy evaluation time-driven.
\subsection{Example: an output statistics}
Let us discuss
the two approaches in the following simple example.
In the wireless simulation, discussed in Section \ref{s:intro},
it is necessary to output, as a function of time, the number of
mobile users that request but do not obtain a connection.
Together with other information,
this quantity is to be
displayed on the PC screen as the simulation progresses.
Thus, 
given a time increment $\Delta t$, 
for each time instance
$t = 0, \Delta t, 2 \Delta t, \dots$, 
the simulation is supposed to report the number of $trying$
users among all the users present at time $t$.

The lazy method is straightforward:
we scan the list of users and
depending on whether a user is $trying$ or non-$trying$ at
time~$t$, increment the corresponding quantity
$trying(t)$ by one or not.
Before each scanning instance~$t$, we reset counter $trying(t)$
to $0$. 
This is obviously a time-driven method.

The anticipatory method is driven by events.
It does not reset or recompute $trying$ at each reporting time instant.
Instead, it maintains the correct value of $trying$ continuously
by updating it
appropriately at every event which may change $trying$.
Such events are: call arrivals to the system, call releases
from the system, and users that obtain a connection after a
period of retrials.
For example, when a user arrives to the system, we would
increment $trying$ by one if the user does not immediately get
connected.

Note that both solutions ``deliver'' the same value of counter $trying$.
This facilitates the development of the simulation program.
In the initial phase, one may program a simpler lazy solution.
Then an anticipatory solution should be programmed.
Both solutions should then be tried in representative
examples, where first we compare values of the results,
and then, after the values are found identical,
we compare the running times.
\subsection{Space sectorization}
An example of the anticipatory data structure is 
space {\em sectorization}.
Billiards simulation in dimensions two and higher share
this method with wireless mobile simulations.
The query, in the case of wireless simulations, may be
to locate all the mobiles that are close to a given point
$(x, y)$ on the plane.
Similar queries exist in the billiards simulation.
Let us consider the wireless simulation case.
There exists a straightforward method to locate all
mobiles which are within a given distance~$r$ from the point
$(x, y)$.
In this method, we would find the position $(x_m , y_m)$ of each
mobile, compute the distance~$r_m$ from $(x_m , y_m)$ to $(x, y)$,
and then select those~$m$ for which $r_m < r$.
In practice, $r$~is small and there are just a few such~$m$, but to
find them we would have to scan all the mobiles.
 
A structure that helps reduce the computation cost is
partitioning the simulation area into smaller sectors.
The geometry of a sector may be different,
a usual choice is a square.
Given such a square $j$, there
is a set of all the mobiles $\{ m \}_j$ that are inside it.
Knowing positions of all the mobiles, we can, of course, find the set
$\{ m \}_j$ using a total scan of all mobiles when a query is offered.
In the anticipatory method, we maintain the sets $\{ m \}_j$
continuously independently of queries.
This entails updating two such sets
each time a mobile crosses a demarcation line between the squares; 
the crossing becomes an extra event to process.

Having invested in this anticipatory structure, we now find
the mobiles $r$-close to a point $(x, y)$ in a different manner.
Namely, we first determine the square $j_0$ in which $(x, y)$
is located, then scan nearby squares $j$ that intersect with any
circle of radius~$r$ with center anywhere in $j_0$, and then
check only the mobiles that belong to sets $\{ m \}_j$.
\subsection{The balance between anticipatory and lazy computations}
The same two query evaluation approaches,
anticipatory and lazy, 
can be considered with respect to
Web search engines.
For many query types,
when a search request is submitted on a 
Web search service, a lazy {\em on line}
evaluation of the request is non-feasible.
Too few users would be willing to wait for
the server to scan the World Wide Web.
The global scanning is being performed but {\em off line}.
Mechanical and manual methods are used
to create structures which contain
answers to various anticipated questions 
and mechanisms exist to quickly retrieve these answers
concentrated at a few data base sites of the 
search server.

The ``economics'' of a Web server query evaluation 
and that in a simulation
are different, though.
In the former, high computational
and manual labor costs are justifiable by 
the need of short response time.
In the latter, an anticipatory method can be only justified
if total computing effort when using the method decreases.

With a lazy evaluation, while
there may be high cost of retrieving the answer when needed,
no resource is spent for anticipating the question.
With an anticipatory evaluation, 
the cost of retrieving the answer may be smaller,
but possibly a high investment may be needed 
for anticipating the question.
Anticipatory
mechanisms should be advanced 
only as far as savings on the retrieving end
of the balance are higher 
than spendings on the investing end.
\subsection{Example: a circuit switched network}
The situation may be illustrated with 
the following example \cite{EGLW93}
of simulating a circuit-switched wired network, 
like the AT\&T telephone network.
In the model,
the network consists of $N$ nodes that represent the switches
and $L = N(N-1)/2$ links that connect the nodes.
A link consists of a fixed number (possibly zero) of {\em trunks}.
The model assumes, that
if a call is to be carried along some path in the network,
then one trunk from each link on the path 
must be allocated for the exclusive use of the call.
The model also assumes that calls are randomly generated
between node pairs ($n_1 , n_2$) and that a {\em routing policy}
decides whether to block (reject) such a call or
carry it on some path between $n_1$ and $n_2$.
The simulation is used to assess the quality of different routing policies
in terms of the blocking produced for given traffic loads.

Consider the 
{\em Least Busy Alternative} (LBA) policy
and
the {\em Aggregated Least Busy Alternative} (ALBA)
policy.
Both policies allocate a trunk on the link 
between nodes $n_1$ and $n_2$ if an idle trunk is available.
If not, both offer the call 
to a two-link path that uses an additional node $\nu$.
The intermediate node $\nu$ is called the {\em via}.
In LBA policy, the overflowing call is offered to
the two-link path that has the most idle capacity.
Specifically, path $(n_1 , \nu , n_2)$ is used
if $\nu^* = \nu$ maximizes
\begin{equation}
\label{lbamin}
\text{min} (idle(n_1,\nu), idle(\nu,n_2))
\end{equation}
where $idle(x,y)$ is the number of idle trunks currently
on link $(x,y)$.
If no idle capacity is available 
- expression \eqref{lbamin} is zero for all $\nu$ -
then the call is blocked.

The ALBA policy is a coarsening of LBA, which
is less costly to implement.
The overflow calls are offered to 
a two-link path with a minimum {\em load class}.
Load classes are defined by a fixed number of prespecified
capacity boundaries. For instance, three load classes
can be defined with boundaries
\mbox{ 0\% $= g_0 < g_1 < g_2 < g_3 =$ 100\% },
where say $g_1 = $80\% and $g_2 =$90\%.
If on a given link the proportion of occupied trunks is $g$,
then the link belongs to the load class $i$ with
$g_{i-1} \le g < g_i$.
Thus, class 1 contains the lightly loaded links,
class 2 contains the moderately loaded links,
and class 3  contains the heavily loaded links.
The load class of a two-link pass is defined to be the smaller
of the load classes of the two links.
An overflowing call 
between $n_1$ and $n_2$ uses a two-link path
whose load class is the minimum among all two-link paths
between this node pair.
If no two-link path has sufficient idle capacity,
then the call is blocked.

The query subject to discussion here 
is finding the via $\nu$ when no idle
capacity exists on the direct path $(n_1 , n_2)$.
The ``lazy'' method operates exactly as the definition says:
when the query is issued, it scans all possible vias $\nu$, 
for each $\nu$ computes the idle capacity
\eqref{lbamin} and chooses the $\nu$ 
which yields the maximum to \eqref{lbamin} in the case of LBA policy,
or it chooses a via with the minimum of the load class 
in the case of ALBA policy.

An anticipatory data structure 
may be suggested for the case of LBA policy
which 
for each
node pair ($n_1 , n_2$)
consists of a list of vias $\nu$ 
sorted decreasingly according to \eqref{lbamin}.
Whenever a least busy via is needed, 
the first via in the list is chosen
which greatly decreases the computations
on the retrieving end of the balance.
On the investing end of the balance,
each time a trunk is allocated or freed
on any link $\nu_1, \nu_2$,
the $(N-1)(N-2)$ sorted lists for all 
links ($n_1 , n_2$) which can
use either $\nu_1$ or $\nu_2$ as a via
have to be updated.
As a result, there is no overall saving
if anticipatory approach is used for the LBA simulation.

However, for the ALBA simulation, the anticipatory approach
yields a significant improvement comparing with the lazy approach.
The capacity of a link is usually measured in hundreds
if not thousands of trunks
while there are typically only a few, say three load classes.
Thus, when a link trunk occupancy changes,
the link rarely changes its load class.
If the load class does not change, no expensive
update of sorted class-via lists is needed.

In the ALBA case, both lazy and anticipatory methods were tried.
While a useful lazy simulation run of the network 
took more than 100 hours,
the corresponding anticipatory simulation run took 3 to 10 hours 
on the same uniprocessor.
\subsection{Anticipatory and lazy billiards simulations}
We now examine event-driven
billiards simulations with respect to
the topic of anticipatory vs. lazy evaluation.
The query concerning a ball location is better accommodated
by the anticipatory sectoring as discussed above.
Another basic query in the billiards simulation is:
which collision has to be processed next?
In the anticipatory method of D.\ Rapaport \cite{R80},
the data structure, which accommodates the answer,
includes for each ball a set of all future events which
can not be easily excluded.
(Future collisions with distant balls 
in dimensions two and higher 
can be excluded using the sectoring method.)
The minimum time event in this list is 
the best candidate event 
to next occur with the given ball.
The minimum time event among all
the best candidates is the query answer.
This event is to be processed next.

Maintaining such lists is 
an easy task for the gutter billiards,
as there are only two candidates for a collision with any ball $i$,
its left neighbor $i-1$ and its right neighbor $i+1$.
(The neighbor may be a gutter end for balls 1 and $N$.)

In higher dimensions, 
the task of maintaining the lists becomes more involved,
since many balls can not be excluded beforehand
as candidates for future collisions of a ball,
even if using the sectoring method. 
These anticipatory lists may include 
large and variable number of candidates
and they
have to be examined and updated for 
each event processed.

On the positive side, method \cite{R80} handles easily
the collision {\em preemption} which occurs as follows.
Suppose 
the algorithm
schedules a collision of two balls, say $A$ and $B$,
for some future time $t_{AB}$ 
and this collision becomes the best candidate 
event to occur for each ball $A$ and $B$.
However,
before the processing reaches time $t_{AB}$, 
a third ball, say $C$,
intercepts ball $A$ with a collision scheduled to occur at time $t_{AC}$
so that $t_{AC} < t_{AB}$.
Now, of course, the collision of $A$ with $C$ becomes
the best candidate for $A$.
What should become the new best candidate for ball $B$ ?
The answer is ready, anticipated in the list of ball $B$: 
the previously rated next-best candidate
event for ball $B$
is upgraded in status to become the best candidate.

This anticipatory event-driven algorithm
can be contrasted
with the method \cite{LS90}, 
which also presents an event-driven,
but ``lazy'' simulation algorithm for billiards.
Here only one future event is kept in each individual ball list.
Maintaining this structure is easy:
when a better candidate 
emerges, it replaces the old one,
which is simply discarded.

However, we should reexamine for the ``lazy'' method 
the described above collision preemption 
situation with balls $A$,$B$, and $C$.
What will be the new candidate event for ball $B$ ?
The solution \cite{LS90} introduces a new event type
on a ball trajectory, an ``advancement'' event.
When processing such an event, the ball is advanced
to the new position without changing its velocity.
Thus, when the collision between $A$ and $B$,
which was scheduled for time $t_{AB}$, is preempted
by an earlier collision between $A$ and $C$,
the new candidate event for ball $B$ is
such an advancement to the former collision site with ball $A$.
The time of the advancement event is $t_{AB}$,
which is the time of the previously scheduled and then discarded
collision of $A$ and $B$.
This is similar to the time-driven $\Delta t$ advancement 
in \eqref{ballstep}; the lazy method \cite{LS90}
can be viewed as a fall back to a time-driven mechanism.
However, the lazy method proved to be
not slower than the anticipatory method,
while the overall algorithm is simpler.
\section{Poisson dispenser}\label{s:poisson}
\subsection{A model of telephone providers competition}
Consider in more detail mentioned in Section \ref{s:intro}
econometric model \cite{LLRM95} of two competing telephone providers,
name them company 1 and company 2
(the model generalizes easily to $M > 2$ providers).
The model also includes $N$ telephone customers.
At any time instant $t$ 
each customer subscribes 
either to company 1 or to company 2.
Let $s_i (t)$ denote the subscription status of customer $i$,
$i = 1,...N$, at time $t$, so
$s_i(t) = 1$ or
$s_i(t) = 2$.
Potential bill amounts $B_{ki} =B_{ki} (t)$ can be computed 
for each customer $i$
if being a subscriber to company $k$, $k=1,2$.
That is, 
the service would cost $B_{ki}$ dollars per month
to customer $i$ 
if $s_i = k$.

For example, to compute $B_{ki}$ under 
the MCI's ``Friends and Family'' plan,
we add up the minutes-per-month customer $i$ calls 
all his/her calling parties
who subscribe
to the same provider (here the MCI), and multiply this
by the plan discount price.
Then we add the minutes-per-month customer $i$ calls 
the calling parties
who subscribe to the other provider,
multiplied by the larger regular price.

According to this model,
a tendency to switch the provider
originates in comparing alternative bills:
if $B_{1i} < B_{2i}$ 
then customer $i$ wants to be served by provider 1,
and if $B_{1i} > B_{2i}$,
then by provider 2.
A customer who subscribes to one provider
but wants to be served by the other one
is subject to a ``pull'' to the opposite provider.
The intensity of the pull toward provider 1 
for customer $i$ who is currently with provider 2
in case 
$B_{1i} < B_{2i}$ 
is expressed
by the rate $r_i = f(B_{2i} - B_{1i})$ 
where $f()$ is an explicitly defined 
monotonically increasing function.
If 
such customer $i$
is with provider 2 at time $t$,
then during a small time interval $(t,t+\Delta t)$,
customer $i$ tosses a coin and
switches to provider 1 with probability
$r_i \Delta t$. 
If the switch attempt is not successful,
the switch is similarly reattempted during the next interval
$(t+\Delta t,t+2\Delta t)$, then next interval and so on. 
A similarly defined ``pull'' 
toward company 2 affects a customer $i$
who is currently served by but is not happy with company 1.
The switch attempts are statistically independent of each other
for different customers within the same $\Delta t$ interval
and they do not depend on the system state before time $t$.

The model assumes an individual calling pattern for each customer $i$;
the pattern is defined by 
specifying minutes-per-month $v_{ij}$ customer $i$ calls
each ``friend'' $j$ of his/her.
The calling habits of the customers are stationary,
that is, the calling volume matrix 
$\{v_{ij} | 1 \le i,j \le N\}$ is independent of time.
The parameters of the providers' plans,
e.g., prices, are also time-independent.
(These assumptions can be relaxed.)
Despite the stationarity,
$B_{ki}$ and the strength $r_i$ of the pull may change with time,
$B_{ki}~=~B_{ki}(t)$, $r_i~=~r_i(t)$.
This is so, because 
in plans like ``Friends and Family'' 
when your call-destination ``friend''
switches the allegiance,
it affects your bill.
You become more susceptible to switching to the same
provider if your ``friend'' has done so,
and that is what plans 
like ``Friends and Family'' count upon.
\subsection{Methods of simulating telephone providers competition}
An obvious method 
of simulating the outlined model is time-driven,
it proceeds exactly as the model states.
Time is increased in small $\Delta t$ steps.
At each step 
each customer who is not satisfied with the provider
randomly attempts to switch.
Then 
based on the new assignments $s_i$,
new $B_{ki}$ and new rates $r_i$
are computed
to be valid for the next $\Delta t$ step.
 
Now we describe 
an alternative event-driven method
of simulating this model. 
The method is based on the observation that
the sequence of times of switching allegiance for each customer
forms a Poisson process with the rate 
$r_i$ which varies in time, $r_i = r_i(t)$.  
Note that a stationary Poisson process 
(that with a constant
arrival rate $r = \text{const}$) 
can be simulated as a sequence
of arrival times 
$t_0 < t_1 < t_2 ...$ with independent
distributed exponentially with mean $1/r$
interarrival times.
That is, given arrival $t_m$, to sample $t_{m+1}$
one draws an independent sample $q$  
of a random value
uniformly 
distributed in $0 < q < 1$, and then one computes
$t_{m+1} = t_m - \frac {\text{log}_e (q)} {r}$.

The Poisson property is preserved when
several arrival processes are 
aggregated into a single arrival process.
Say, we have $N$ Poisson arrival processes and $i$th process
has arrival rate $r_i (t)$, $i=1,...N$.
The aggregate process is defined as the one that has
an arrival when any component process does.
So defined, the aggregate is also a Poisson process.
The arrival rate of the aggregate is 
$R(t) = \sum_{i=1}^N r_i (t)$.

One way of arranging 
an event-driven simulation of the competition 
between the telephone providers
is as follows.
A next anticipated switch event
is associated
with
each of the $N$ customers.
The event time is generated by sampling the exponential distribution
with the current rate of ``pull'' discussed above. 
The case of rate 0 is reserved for customers who are satisfied
with the provider.
Such a customer does not wish to switch and
the next switch time is set to infinity.
The event queue is arranged as usual. 
Processing each customer switch modifies some rates $r_i$. 
This, in turn, modifies the time remaining to switch 
in the events scheduled for the affected customers;
sometimes it may reduce the rate to 0
which would postpone the
switch event to the infinite future.
 
Having the Poisson property preserved under the aggregation,
the event-driven simulation of this model can be arranged
in a different way
without presampling and then updating future switch events 
and without the event queue.
In the alternative method, these are replaced with
the following {\em Poisson dispenser} mechanism.
A single Poisson arrival stream with rate $R(t)$ is generated
and then is being ``dispensed'' among the component Poisson streams
in accordance with their partial arrival rates $r_i(t)$,
larger the $r_i(t)$ more probable 
is to delegate an arrival of the aggregate process
to customer $i$'s process.

Rate $r_i(t)$ of a component Poisson process
may experience changes within the time interval 
between the consecutive arrivals of the component process.
By contrast, the aggregate rate $R(t)$ remains 
constant between arrivals of its process.
Hence the interarrivals of the aggregate are distributed
exponentially; 
once next event is scheduled its time never gets changed
or postponed.
This simplifies and speeds up the computing.
The dispenser procedure consists of the cyclic repetition of the 
two steps specified below; 
the execution begins with $m=1$ 
and current time $t_1 = 0$.
\\
\\
\fbox{
\begin{minipage} {12.3cm}
DO
\begin{enumerate}
\item 
Using the current aggregate arrival 
rate $R(t_m) = \sum_{i=1}^N r_i(t_m)$,
\\
sample 
the time increment $\Delta t_m$ 
from
the current time $t_m$
\\
to
the next arrival 
of the aggregate process.
\item Select a component process $i$ with probability 
$p_i = r_i (t_m) /R(t_m)$.
Advance the current time to $t_{m+1} = t_m + \Delta t_m$ 
and change the state of component $i$.
Set new arrival rate $r_i$ for component $i$ and
compute new arrival rates $r_j (t_{m+1})$ for components $j$
whose event arrival rates
may change as a result of the state change of component $i$.
\end{enumerate}
UNTIL the simulation is complete
\end{minipage}}
\\
\\
\\
This procedure is formulated above for a general multicomponent system.
In the considered example, a telephone customer is a component,
and the state change of component $i$ is switch 
$s_i(t_{m+1}) = 3 - s_i(t_m)$ 
of the telephone provider for customer $i$, 
where
map $x \leftarrow  3-x$ turns 1 into 2 and 2 into 1 
as required to effect the provider switch.
Since after the switch the customer is ``happy''
with the new provider,
the new arrival rate in Step 2 is zero;
it may become positive again later 
as a result of ``friends'' switching.
Also, 
values
$p_i$, 
are indeed a probability distribution,
as
$p_i \ge 0$
and $\sum_{i=1}^N p_i = 1$.

In a straightforward method
the selection in step 2
is implemented as follows.
First, we draw an independent sample $q$  of a random value
uniformly distributed in $0 < q < 1$. 
Then we scan monotonically non-decreasing 
sequence $V_0 = 0$,
$V_1 = r_1 (t_m)$,...
$V_i = \sum_{j=1}^i r_j (t_m)$,...
$V_N = \sum_{j=1}^N r_j (t_m)$,
fitting value $Rq$ between consecutive terms
$V_{i-1} \le Rq < V_i$.
This is possible, since $0 = V_0 < Rq < V_N = R$.
The found $i$ is index of the customer to whom the arrival is delegated.
Unfortunately, 
an order of $N$ computations 
is required 
for the scanning and
this is slow for a large $N$.

\begin{figure}[h]
\centerline{\psfig{file=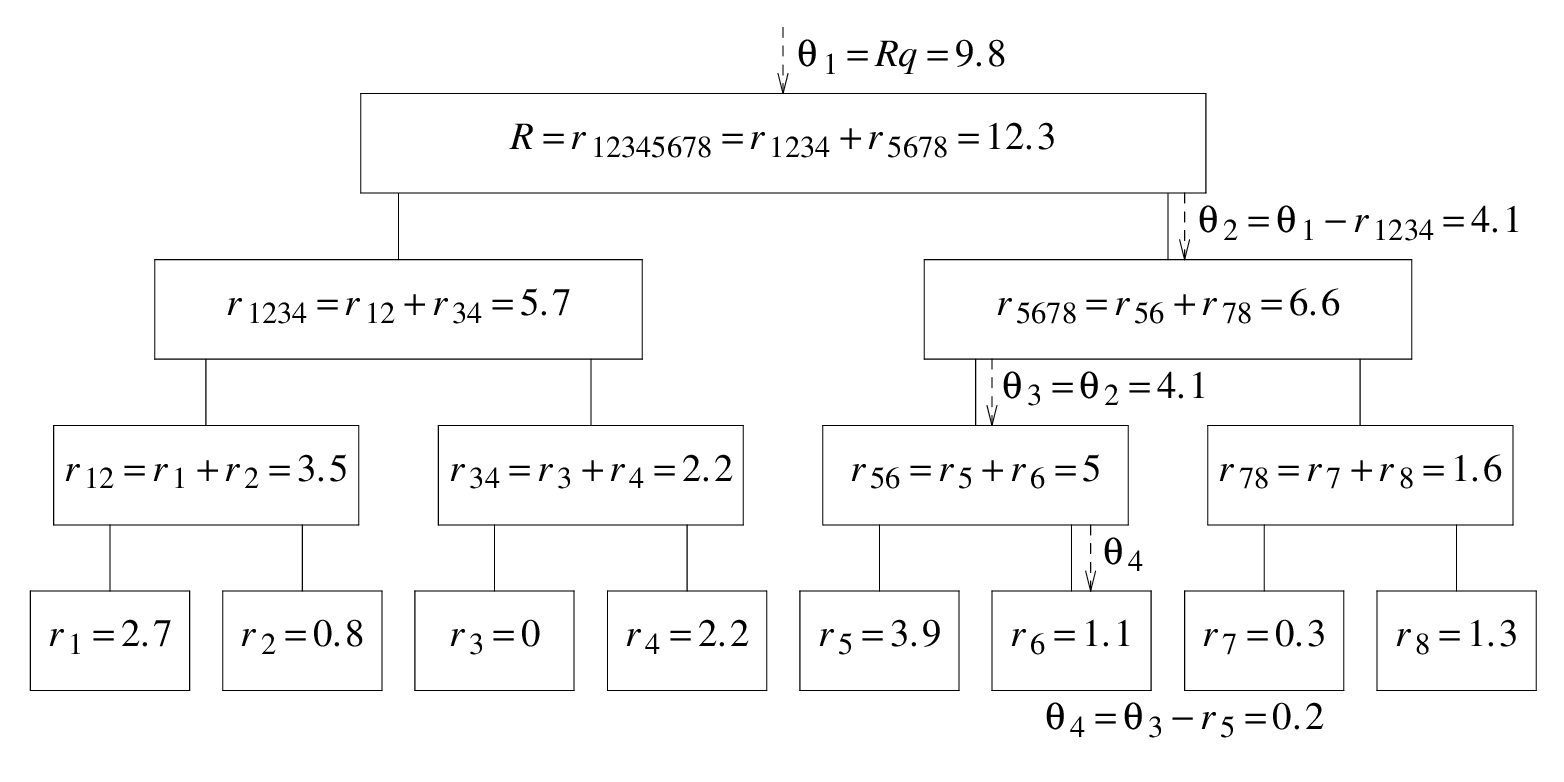,width=6.1in}}
\caption{Tracing down a binary dispenser tree 
with $N=8$ terminal nodes.
An arrival of the aggregate process
is delegated to
a component process in $\text{log}_2 (N) = 3$ steps.}
\label{tree}
\end{figure}

Better methods exist.
Figure \ref{tree} exemplifies the binary tree method
in which
the component $i$ 
is found in $\text {log}_2 (N)$ steps
(if $N$ is a whole degree of 2; otherwise 
$\lceil \text {log}_2 (N) \rceil$ steps).
This method also starts with
an independent sample $q$  of a random value
uniformly distributed in $0 < q < 1$.
Then, instead of a linear scan,
\marginpar{\em Figure \ref{tree}}
a binary tree is traced
to fit this value.

The tracing is started from the root 
which is entered with $\theta_1 = Rq$.
We then steer our way down the tree in $\text {log}_2 (N)$ steps.
At step $m+1$, 
having value $\theta_m$,
$0 \le \theta_m < r_{1,..j,j+1,..k}$,
we enter node with inscription
$r_{i,..j,j+1,..k} = r_{i..j} + r_{j+1,..k}$ on it.
If $\theta_m$ happens to be smaller than $r_{1,..j}$, 
then we go to the left branch $r_{1,..j}$ 
producing $\theta_{m+1} = \theta_m$.
If $r_{1,..j} \le \theta_m < r_{i,..j,j+1,..k}$, 
then we go to the right branch $r_{j+1,..k}$ 
producing $\theta_{m+1} = \theta_m - r_{1,...j}$.

To maintain current the weights on the tree,
each time a rate $r_i$ changes,
the contents of $\text{log}_2 (N) + 1$ nodes is updated.
The update begins with the bottom node $i$, continues up,
and terminates at the root.

A useful simulation run is the switching history
of several thousand customers during a few simulated years.
A time-driven version takes several computing hours to complete the run.
The corresponding  event-driven version with a Poisson dispenser
implemented using the binary tree, takes several seconds.
The latter version allows one to easily simulate markets
of a much larger size
while keeping the running time
bearable. 
For example the behavior of 
a 100,000-customer market
during several simulated years 
can be simulated in less than 2 minutes.
\subsection{Ising spin simulations}\label{ss:ising}
Another application for the dispenser technique,
is an {\em Ising spin model} \cite {I25} \cite {BKL75} 
in computational physics.
As a computational mechanism,
the Ising spin simulation is very similar
to the econometric model of competition between the telephone
providers which has been just discussed.
$N$ ferromagnetic
particles are
in place of $N$ telephone customers.
An external magnetic field 
pulling the particles so that 
they would align in the field's direction
plays the role
of the
economic incentive
for customers to ``align'' their allegiance to the provider
which gives best saving.
And the additional local magnetic field around a particle
when its neighbors have already aligned is analogous
to the additional incentive 
for a customer to switch the provider
when the customer's calling parties have done so.

Specifically, the magnetic state 
called {\em spin}
of each particle $i$,
takes on two values,
$s_i (t) = 1$
or
$s_i (t) = -1$.
Depending on the current spins $s_j (t)$
of the 
near-neighbors $j$ of particle $i$,
pulls to flip $s_i(t) = 1$ to $-1$ and 
to flip $s_i(t) = -1$ to 1 are defined.
The dependence
involves external field direction and intensity.
(The form of this dependence
is not essential for the present discussion.)
Similarly
to the model discussed before,
the flip mechanism is probabilistic.
Simulation of the dynamics of Ising spins can thus be arranged
using the efficient binary tree Poisson dispenser mechanism
discussed above.
In this method flipping one spin 
takes $\text{log}_2 (N)$ computations.

One feature in the Ising model is different 
from the telephone competition model though. 
The Ising particles are arranged in a regular fashion.
For example, they are placed in vertices 
of a two-dimensional square lattice where
each particle has four near-neighbors:
at the North, East, South, and West.
By contrast, the calling volume matrix $v_{ij}$
which defines calling parties of a customer
is not assumed to be regular.
Because of the regularity in the Ising model, 
for any particle,
there is only a small number
of possible combinations of neighboring particle states.
Let us count these combinations, say, 
for a planar square grid arrangement
of the particles.
Each of the 
North, East, South, and West spin can take on 2 values,
which yields
16 combinations.
The particle itself can be in the two spin states.
Hence there are at most 32 different combinations
each of which may define a different
pull-to-flip rate.
Therefore there are at most 32 different rates
and this finite set is the same for all $N$ particles.
(Examining the form in which the rate depends on the combination,
this set further reduces to
only 10 different rates \cite{BKL75}.)
Thus, each of the $N$ leaves
in the dispenser tree 
contains one of a small number of possible values.
The finiteness allows one \cite{BKL75}
to improve the Poisson dispenser algorithm
so that delegating
one arrival of the aggregate process to a component process
takes a constant amount of computations, 
instead of 
$\text{log}_2 (N)$ computations.
\section{Simulating sequential random update in parallel}\label{s:sequent}
Sequential random update is a general mechanism for modeling
evolution of a multicomponent dynamic system.
In this model, we consider a 
system with $N$ components, 
the state of the system 
is composed of states $s_i$ of its components,
The system state changes at discrete instances
$m = 1, 2,....$ according to the following cyclic procedure.
\\
\\
\fbox{
\begin{minipage} {12.3cm}
DO
\begin{enumerate}
\item Select a component $i$ randomly and uniformly in the range 
\\
$1 \le i \le N$.
\item \label{stateupd} Change the state $s_i$ of the selected component.
The new state $s_i^{m+1}$ is a function of the old state 
$s_i^m$ and perhaps the states of some other components $s_j^m$.
\item Increment $m$ by 1.
\end{enumerate}
UNTIL enough cycles are processed
\end{minipage}}
\\
\\
\\
Both the Ising spin model and
the model of competition between telephone providers 
discussed in Section \ref{s:poisson} fit
the sequential random update scheme,
if two transformations are made:
1) uniformizing the event arrival rates 
among the system components,
2) abandoning the continuous time
and retaining only the update counter $m$.

The latter transformation is obvious.
Let us explain the former one.
We choose an upper bound $r_*$ on the event arrival 
rates $r_i(t)$ among the components.
This can be easily done in the Ising model:
$r_*$ is the largest among the
finite number of possible rates $r_i$,
see the discussion in Section \ref{ss:ising}.
In the provider competition model,
the most ``unhappy'' customer yields
the upper bound on the switch rate.

We attribute the same rate $r_*$ to all components so that
the components are selected with equal probability
as required in Step 1 above.
Suppose a component $i$ at Step 1 is selected.
To compensate for the smaller rate 
with which the component $i$
is being updated,
we make an additional coin toss and choose
to update state $s_i$ in Step 2 with probability
$r_i/r_*$. 
With the complimentary probability $1 - r_i/r_*$
the state does not change.
The no-change of
state $s_i$
does not violate the format
of the procedure; it is a special case of the update when
$s_i^{m+1} = s_i^m$.

The model of circuit-switched wired network 
discussed in Section \ref{s:lazy} almost fits this scheme,
if we take $N(N-1)/2$ links between the switches as components
and equate the number of occupied trunks on a link
with the state of the component.
If different node pairs $(n_1 ,n_2)$ have different call arrival
rates, we uniformize the rates as in the previous two models.
The feature that does not fit the scheme is 
that when a call request arrives
between a node pair $(n_1,n_2)$ and is placed via node $\nu$ instead
of the direct link,
the state of components $(n_1,\nu)$ and $(\nu,n_2)$  
is updated
instead
of the state of the original component
$(n_1,n_2)$.
\subsection{Ballistic particle deposition}
\begin{figure}[h]
\centerline{\psfig{file=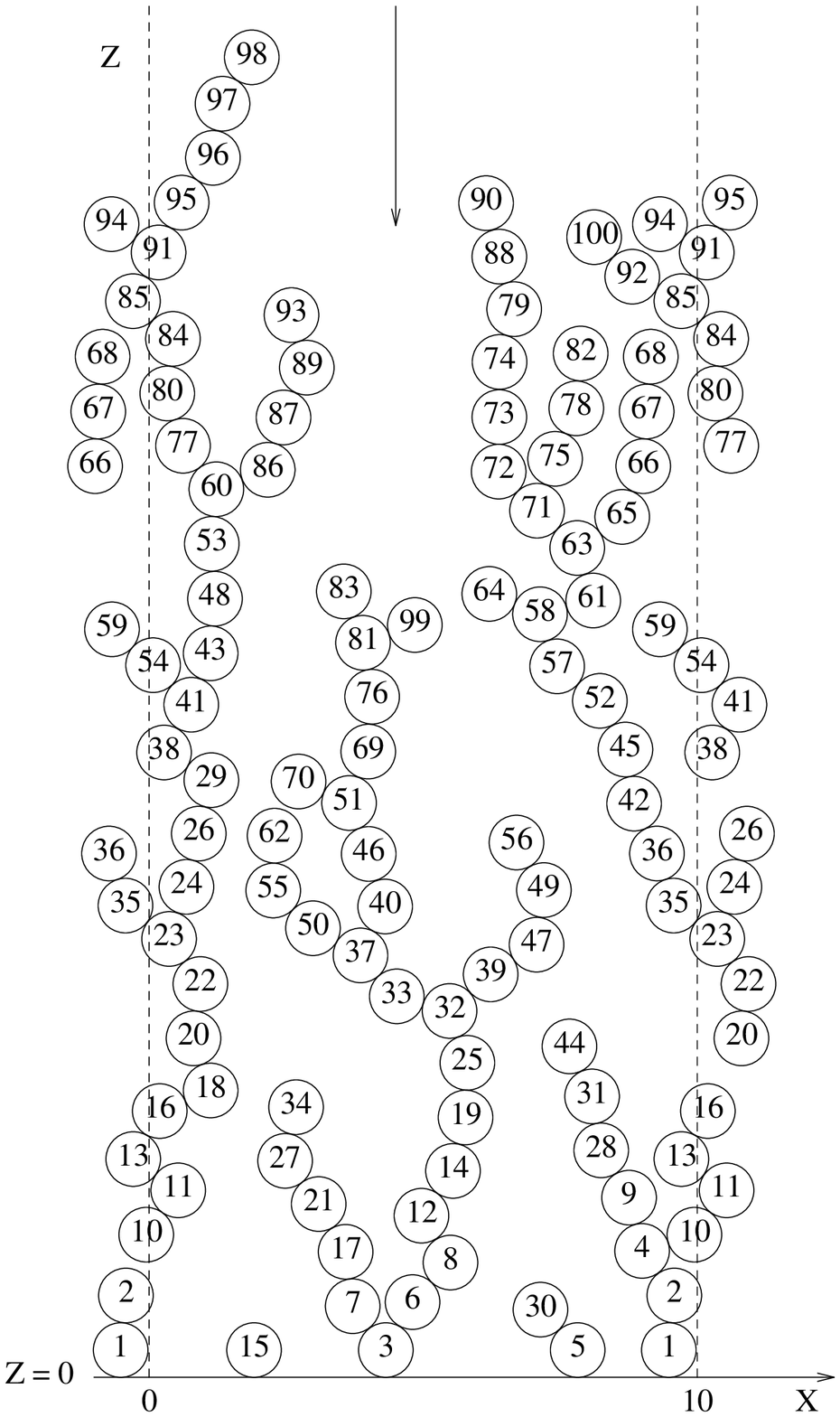,width=4.0in}}
\caption{First 100 particles deposited over a segment of length 10.
Particle 1 was deposited first,
then particle 2, and so on. 
The endpoints of the segment are ``glued''
together to form a circle. 
Because of this,
the particles that fell close to a dashed boundary, 
like particle 1, are shown twice.}
\label{100disk}
\end{figure}

Other instances  of the sequential random
update include cellular arrays and neural nets.
We will now discuss an example \cite{LPR96} 
of a {\em ballistic particle deposition}.
The deposition model is aimed at 
studying 
morphology of amorphous layers growing on planar substrates,
the subject of interest to material scientists.
In the model,
spheres of equal diameters 1 are falling vertically down
toward the flat
\marginpar{\em Figure \ref{100disk}}
horizontal surface,
see an example in Figure \ref{100disk}.

The deposition process is arranged as follows.
Particles are deposited one at a time.
To deposit a particle, first
its center $x$ coordinate is sampled randomly and uniformly
over the adsorbing range.
The range is segment (0,10) in Figure \ref{100disk}.
Different particles have their $x$ 
generated independently of each other.
The initial height of the center,
is chosen sufficiently high above
the surface.
Then the particle is falling vertically down
until a contact occurs.
For the particles that begin the process,
the contact is likely to be with the adsorbing surface.
Later in the process, a falling particle is more likely
to contact a stationary one,
deposited before.
In either case, the first contact 
instantly stops the incoming particle
thus defining its $z$ coordinate.

The ballistic
deposition scheme is
an instance of the random sequential update.
We split 
the adsorbing range into 
sectors of equal measure 
(congruent segments in 2D or planar figures
of equal area in 3D)
and declare sectors to be 
the components of the system.
The particle order index $m$ becomes the update order index.
The state $s_i^m$ of component $i$
consists of the coordinates of those among the first $m$ particles,
that have been deposited over sector $i$.
That is, say in a two dimensional case,
the $x$ of those particles centers must be in segment $i$.

A particle from
sector $i$ but located close
to the sector's boundary may attach itself to a particle
which was earlier deposited 
in sector $j$, $j \ne i$.
That is how components/sectors $j$ may get involved in
the state update of component $i$ in 
Step \ref{stateupd} of the sequential update scheme.

\begin{figure}[h]
\centerline{\psfig{file=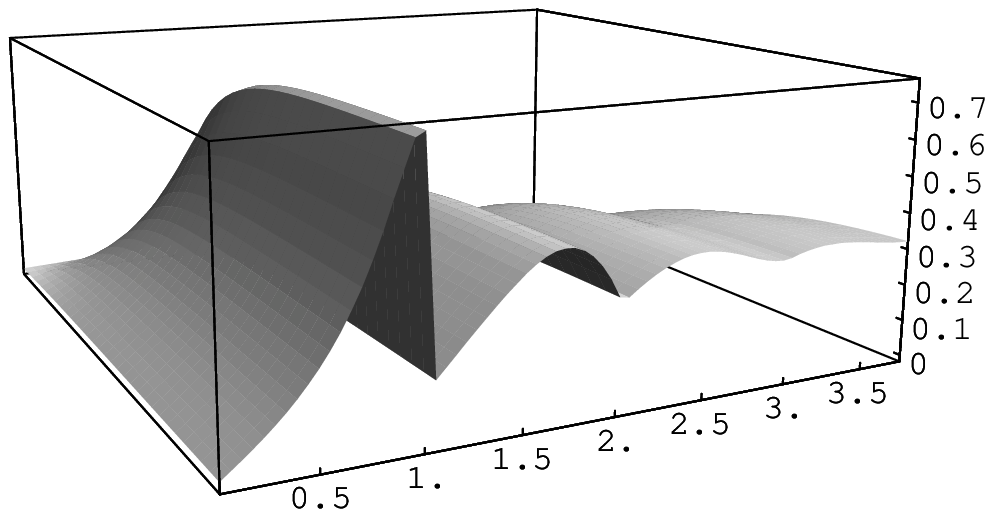,width=6.0in}}
\caption{The density of deposited particles as a function
of height and deposition time.
The height, measured as $z-1/2$,
is changing along the lower front
horizontal edge of the box, 
the time is changing along the lower left horizontal edge.
The rear facet of the box is the plane of zero time.
The density is changing vertically.}
\label{cumden}
\end{figure}

Whereas Figure \ref{100disk} represents 
a small-scale 
``educational''
example of deposition in 
two dimensions,
interesting simulations are in three dimensions
with two dimensional adsorbing range of
a size larger than say 1000$\times$1000. 
Many millions if not billions of particles
are supposed to be processed.
\marginpar{\em Figure \ref{cumden}}
Figure \ref{cumden} presents particle density 
as a function of both time and height $z$ 
in a three dimensional
deposition of
100 million particles.
Running such an experiment on a workstation
would take about a week of execution time.
Can parallel processing be employed 
to speed up the deposition and other
sequential update schemes?

\subsection{Methods of parallelizing sequential random update}
One idea of making the update scheme parallel may be to 
have a parallel computer dedicate $N$ processing elements (PEs)
to the $N$ components of the system so that $\text{PE}_i$ would host
component $i$, $i = 1,2,...N$. 
The components would
be updated concurrently 
without an organizing order.
$\text{PE}_i$ would repeatedly update the state of component $i$,
asynchronously obtaining from other PEs
the current values of states of those components $j$
which are 
required 
for computing the new value of state
of component $i$.

This proposal can be criticized from several viewpoints.
The most basic objection is that it violates 
the standard model of reproducible computer execution.
This entails various shortcomings.
The generated trajectory of the system may be different
from that generated sequentially. 
Say in the deposition example, 
the deposit structures generated in parallel 
may be statistically different
from those generated sequentially.
Moreover,
the state change trajectories resulted
in different runs of the same system with the same
initial states, in general, will be different.
This is very inconvenient as it makes both
studying the obtained structures 
and debugging the simulation program 
very difficult.

Another proposal is to use $N+1$ PEs.
A single ``master'' PE would be
dedicated to 
dispensing the updates among the components.
The $N$ ``slave'' PEs
would host 
the $N$ components.
A ``slave'' would be responsible 
for updating the state of the hosted component.
Without further elaboration of the ``master-slaves'' scheme,
we note that it can be organized so that the computations
are reproducible and generate trajectories identical
to those in the sequential procedure.
Moreover, for a small number of PEs,
the procedure may even be efficient.
However, it does not scale for a large number of PEs.
The sequential dispensing performed by the ``master''
becomes a bottleneck for a large $N$.
\subsection{Cautious advancement method}
In the example of ballistic deposition, we now 
describe another method 
of running the sequential random update scheme in parallel.
Unlike the ``chaotic'' and the ``master-slaves'' methods,
this method possesses both desirable properties:
it generates a reproducible, correct simulated trajectory,
and its good performance scales to a large size systems 
and large number $N$ of PEs.

The first step of the method is a reformulation
of the sequential random update scheme 
in continuous time. 
In the old formulation the components are updated
at discrete instances $m=1,2...$.
In the new formulation the components are updated
at instances $t_1 ,t_2 ,...t_m,...$ of the continuous time,
the instances constitute a Poisson process.
An arbitrary positive $\lambda$ is chosen and fixed.
The rate of the Poisson process is $N\lambda$.
It follows, that each of the $N$ component processes
also has arrivals that form a Poisson process.
The component processes will be mutually independent.
The rate of each component process will be $\lambda$.

Of course, in the Ising spins simulation model
and in the provider competition model we have the
Poisson processes to start with. 
We can reuse them
with their original rates
for the purpose of rendering the models in parallel.
However, in the deposition model,
the Poisson process is an additional structure,
introduced only for the purpose
of running the model in parallel.

In the Poisson dispenser method 
discussed in Section \ref{s:poisson},
we aggregated
arrivals of individual components into a single stream
which was to be sampled by the 
sequential computer.
Here we do the opposite, namely, disaggregate
the summary Poisson arrival stream into the
independent streams of the components 
and let each component $i$ stream be sampled by a separate 
$\text{PE}_i$ which would be also responsible
for maintaining state $s_i$ of component $i$.

Let $t^i$ denote the Poisson clock maintained by 
$\text{PE}_i$. 
That is, changes of state $s_i$
occur at time instances 
$t_1^i , t_2^i ,....$.
In the beginning of simulation
each 
$\text{PE}_i$
sets its Poisson clock $t^i$ to 0
and has $s_i$ assume its initial value.
Then each
$\text{PE}_i$
asynchronously from other PEs executes the following procedure,
which can be called ``cautious advancements.''
\\
\\
\fbox{
\begin{minipage} {12.3cm}
DO
\begin{enumerate}
\item Sample next arrival $t_i$ of the Poisson process with rate
$\lambda$.
\item \label{wait} If changing state $s_i$ at time $t_i$ requires
the values $s_j$ of states of other components $j$, then
wait until each such component $j$ reaches time $t_j$
so that it will become $t_j \ge t_i$.
\item \label{pstateupd} Change the state $s_i$ of the hosted components
as required, 
possibly using current values of states $s_j$ of other
components $j$.
\end{enumerate}
UNTIL the simulation of component $i$ is completed
\end{minipage}}
\\
\\
\\
The PEs may execute this procedure 
with no other synchronization 
than that in the wait statement in Step \ref{wait},
which is supposed to assure the ``cautious advancement''
of local times $t_i$.
Because of the asynchrony,
it might be not obvious that the procedure is correctly defined,
let alone works correctly.
For example,
if concurrently with
$\text{PE}_i$
updating $s_i$ and using state values $s_j$,
$\text{PE}_j$ is changing $s_j$,
is the state update in Step \ref{pstateupd} well defined?

It was shown \cite{L87}
that the cautious advancement scheme is correctly defined, 
works and is correct. 
Specifically, with probability 1,
during an update of state $s_i$, other states $s_j$,
the values of which are used in the $s_i$ update,
are not themselves being changed.
Cautious advancement is deadlock-free:
no PE waits forever in Step \ref{wait}.
The sequence of updates obtained by merging
the sequences of component updates generated by each PE
is statistically identical to the sequence 
generated by the sequential procedure.

Reproducibility also takes place:
if the cautious advancement procedure is executed twice 
with the same initial settings,
including the same seeds of pseudorandom number generators,
used for sampling Poisson arrivals,
then
the two resulting sequences of state updates will be identical
with probability 1.
This is despite that execution
timings in different runs may be different.

\begin{figure}[h]
\centerline{\psfig{file=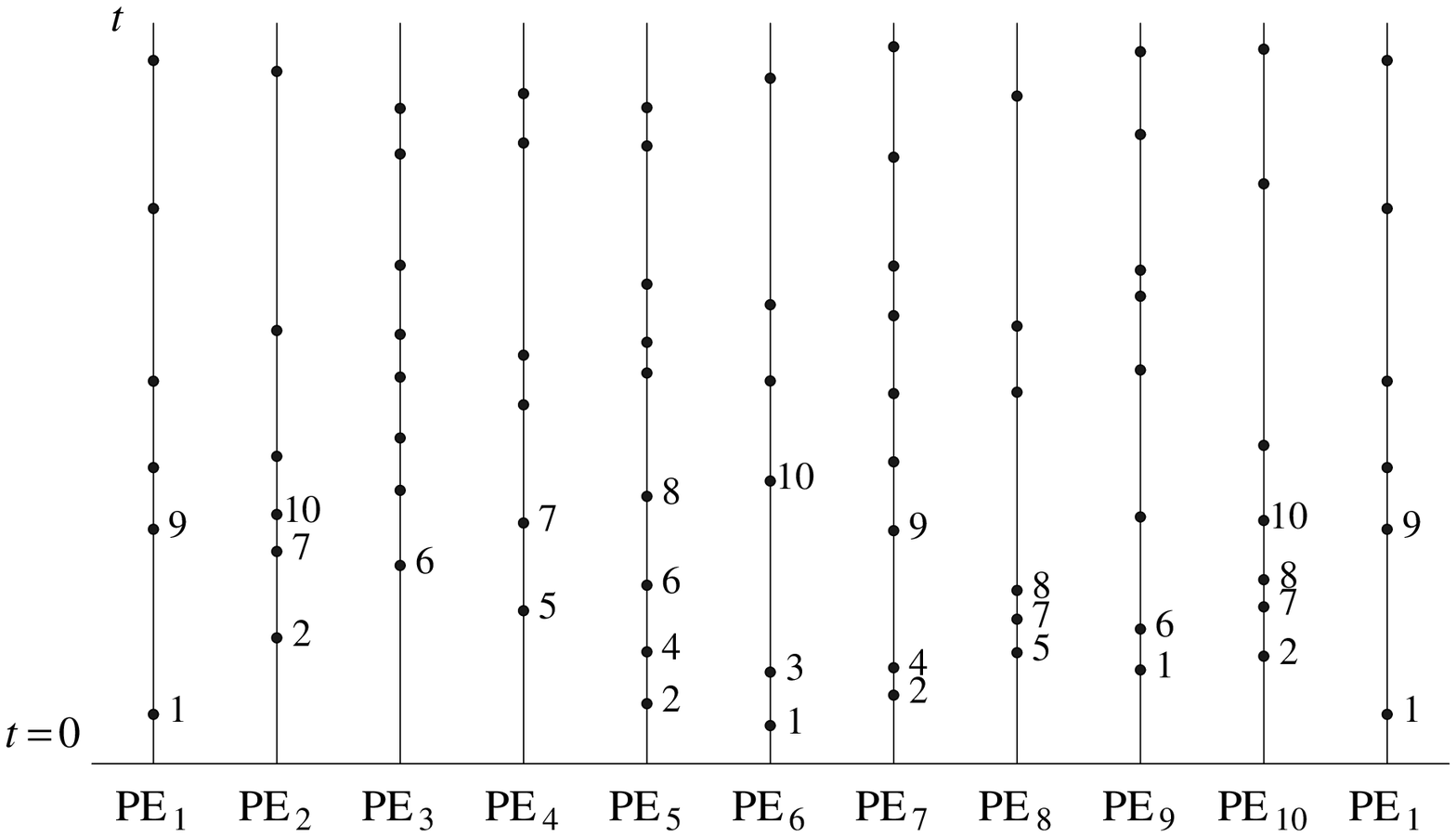,width=7.0in}}
\caption{Local time lines of 10 processing elements
that collectively perform a sequential random update,
e.g., the particle deposition as in Figure \ref{100disk}.
Dots mark the state update instants,
e.g., instants when particles are deposited.
The indices near the dots 
indicate the cycles when the updates
occur}
\label{tlines}
\end{figure}

To get insight into the behavior of the cautious advancements,
consider a simple example presented in Figure \ref{tlines}.
\marginpar{\em Figure \ref{tlines}}
Time lines of 10 components of a simulated
model are depicted.
For concreteness, we assume that 
the model represents the deposition of unit diameter
circular particles over the segment of length 10 
as in Figure \ref{100disk}.
In the latter case, 
the substrate segment (0,10) is divided into 10 smaller segments:
[0,1),[1,2),...[9,10).
Segment [$i-1,i$) is component $i$
and 
it is hosted by $\text{PE}_i$.

Dots on the time lines mark
the Poisson arrivals.
At each arrival to segment
[$i-1,i$), a particle with coordinate $x$, 
$i-1 \le x < i$,
is deposited.
Given that both the particle diameter 
and component segment length are unity,
to know the landing height of the particle,
there is no need to know 
positions of particles 
previously deposited over segments
that are more than one component-segment away.
However, a caution is exercised with respect to 
the immediate left and right
neighbors of segment [$i-1,i$).
$\text{PE}_i$ only 
deposits
a particle at time $t_i$ 
if its two neighbors advanced their simulated times
to reach or exceed $t_i$.

The immediate left neighbor is segment [$i-2,i-1$), unless $i = 1$.
The immediate right neighbor is segment [$i,i+1$), unless $i = N$.
Points 0 and 10 representing the same point,
the immediate left neighbor of component 1 is component $N$, 
whose immediate right neighbor is component 1.
To be able to relate the arrival times for neighbors 1 and $N$,
the time line of component 1 is drawn twice.

Although no additional synchronization 
is necessary for correctness and efficiency 
of the general cautious advancement
procedure,
it would help in understanding the procedure of deposition,
if we assume that it is executed in lockstep.
That is, no PE executes Step 2, before Step 1 is completed
by all the PEs that are non-waiting at 
Step \ref{wait} of
a preceding cycle.
Then,
in Step \ref{wait},
all PEs check the simulated times achieved by their neighbors.
As a result, the set of all PEs splits into 
those able to proceed further,
and the rest which must wait.
The non-waiting PEs 
begin Step \ref{pstateupd} only after
all PEs have finished the checking in Step \ref{wait}.
Finally, the new cycle by the non-waiting PEs
begins not earlier, than all the non-waiting PEs 
have updated their state, i.e., deposited a particle.

The lockstep execution enables us
to say at which cycle each state update occurs,
that is, each particle gets deposited.
In the situation of Figure \ref{tlines},
$\text{PE}_1$,
$\text{PE}_6$,
and
$\text{PE}_9$
are lucky to process
an event at cycle 1, while the rest of PEs are waiting.
Values $t_1, t_6$, and $t_9$ get advanced 
to the second arrival time and as a result
$\text{PE}_2$,
$\text{PE}_5$,
$\text{PE}_7$,
and
$\text{PE}_{10}$
can process their events at cycle 2.
Values $t_2, t_5 ,t_7$, and $t_{10}$ get advanced 
to the second arrival time and as a result
$\text{PE}_6$ is lucky again and processes its second event.

In Figure \ref{tlines},
the cycle-per-event assignment
is followed up to cycle 10.
As an exercise, 
the reader may continue the task for the following cycles.
The computational efficiency is determined
by the fraction of 
non-waiting PEs at each cycle. 
This fraction is close to 25\% on average
in the example shown.
Looking at the picture it seems likely that the fraction
remains bounded from below and separated from zero
when the size of the system gets larger and number of PEs
increases proportionally.
Mathematical studies \cite{GreenSS} of this assertion confirm it
in a general case.
This means scalability of the parallel simulation
performed by the cautious advancement mechanism.
Using additional 
methods \cite {LPR96}, 
in particular, allowing one PE to hold a larger
sector or segment, the efficiency can be substantially increased,
e.g., from 25\% to 60\% and higher.
The mentioned above 100 million particle deposition
experiment was run on a Maspar MP-1216 computer
with 16,384 PEs. 
The run took 620 seconds (instead of a week on a workstation).

It becomes clear from the deposition example, that
the efficiency of 
the cautious advancement parallel method for sequential
random update depends on
the topology of the connections among the components.
The sparse fixed connections and a large
diameter of the connection graph increase the efficiency.
A small-diameter graph with
all-to-all connections or close to such makes
the PEs to be too cautious; too few of them would be
non-waiting.
In the worst case only one PE dares
to advance the local time during a cycle 
while 
all the other PEs
are cautiously waiting.

That is the case in the circuit-switched network simulation,
where node pairs $(n_1 ,n_2)$ 
are close to each other in the sense of
network connectivity.
Even if we somehow resolved the difficulty that this model
does not completely fit the random sequential update model
as discussed in the beginning of the section,
its parallel execution by the described method would not be efficient.
However, the actual event dependency along the executional
path is rather sparse which presents an opportunity for parallelism.
Unfortunately, this parallelism
is not extracted by the cautious advancement method,
because the method requires a variable event dependency graph 
to be upper bounded by the fixed component connectivity graph.

Ising model and the phone provider competition
model have a sparse component connectivity,
but they still may fail to produce an efficient simulation
using the described technique.
This is because, for example
in the Ising model,
among the parameters the spin flip rate depends on
is the temperature \cite{I25},
and a low temperature causes
the ratio between a particle update rate and its upper bound
to be large.
Even though a high enough fraction of PEs 
do not wait with processing their events,
most of the processed events are trivial 
time advancements without a flip.
This ruins the efficiency 
of the parallel execution in low temperature regimes.

A successful use of the cautious advancement technique 
and its further developments for an Ising model
at a non-very-low temperature has been demonstrated \cite{K99}
with the model running on 400 PEs of 
a parallel supercomputer T3E and
yielding a speedup of 260.
Another example is a wireless simulation \cite{GLNW95}
where the calls are dynamically arriving at 
random positions in the service
area according to a fixed distribution
and the users are not moving during the calls.
With such assumptions it is possible to arrange
a variant of efficient cautious advancement parallel processing.

Next section discusses an alternative method of parallel execution.
It aims to extract parallelism as it emerges 
during the execution rather than relying on a worst case estimate
given by the connectivity graph.
Also it needs no uniformization of the event arrival rates.
\section{Synchronous relaxation}\label{s:relax}
As in Section \ref{s:sequent},
we consider a discrete event simulation 
of a dynamic system with $N$ components. 
The simulation is to be
performed on
a parallel computer with $N$ processing elements.
As before, the procedure is to give
each PE a component to host,
and have the PE produce the state change history of
that component.
The method to do so will be different from the one discussed
in Section \ref{s:sequent}. 
Unlike the previous method in which the event processing is final,
the present method involves speculative computations
wherein 
an event can be processed and then later rejected.
The procedure was called {\em synchronous relaxation}\cite{EGLW93}.

In this procedure,
each PE keeps track of
the simulated time before which no
event is to be rejected
in the course of further processing;
this quantity is called {\em committed time.}
The PEs increase the
committed time in lockstep,
its value is common to all PEs.
Each step consists of several iterations.
At each iteration,
while the committed time value does not change,
each PE produces a speculative state change trajectory
of the component it hosts
beyond the committed time.
The PE extends the trajectory 
until
its local time reaches the committed time plus $\Delta t$,
where $\Delta t$ is the step size of
committed time increases.
Unlike the $\Delta t$ of a time-driven simulation
discussed in Section \ref{s:edr},
here $\Delta t$ does not define the accuracy of simulation
and may be not small.

Since components are, in general, connected,
to produce a correct trajectory of its component,
a PE needs to know the correct histories of other components.
But they are not known, because
other PEs are in the same quandary.
The mechanism of generating the correct trajectories is by iterations.
During the first iteration,
each PE makes the simplest assumption about
the trajectories of the other components,
for example, 
that the other trajectories are empty of events,
i.e., states do not change.
This will enable the PE to produce the hosted component trajectory.

After every PE generates the speculative trajectory
for additional $\Delta t$ units,
they compare the trajectories.
This comparison phase is started 
only after all PEs have 
generated the trajectories.
As a rule, a PE will detect inconsistencies between
the assumed and actually generated trajectories
of other components.
If so, PEs perform more iterations.

During subsequent iterations,
if a PE needs to know the trajectory
of a component hosted by another PE, it uses the trajectory
generated in the last iteration.
The goal of producing correct
trajectories at a step is achieved
if no PE detects any inconsistencies
between
the assumed and actually generated trajectories of other PEs.
Once this happens,
all PEs increase committed time by $\Delta t$
and continue.

The synchronous relaxation parallel algorithm \cite{EGLW93}, 
used on a Maspar computer with 16,384 PEs,
cuts the running time 
of simulating a circuit-switched wired network 
to a few minutes
(from several hours in the best sequential
implementation on a fast workstation).

\begin{figure}[h]
\centerline{\psfig{file=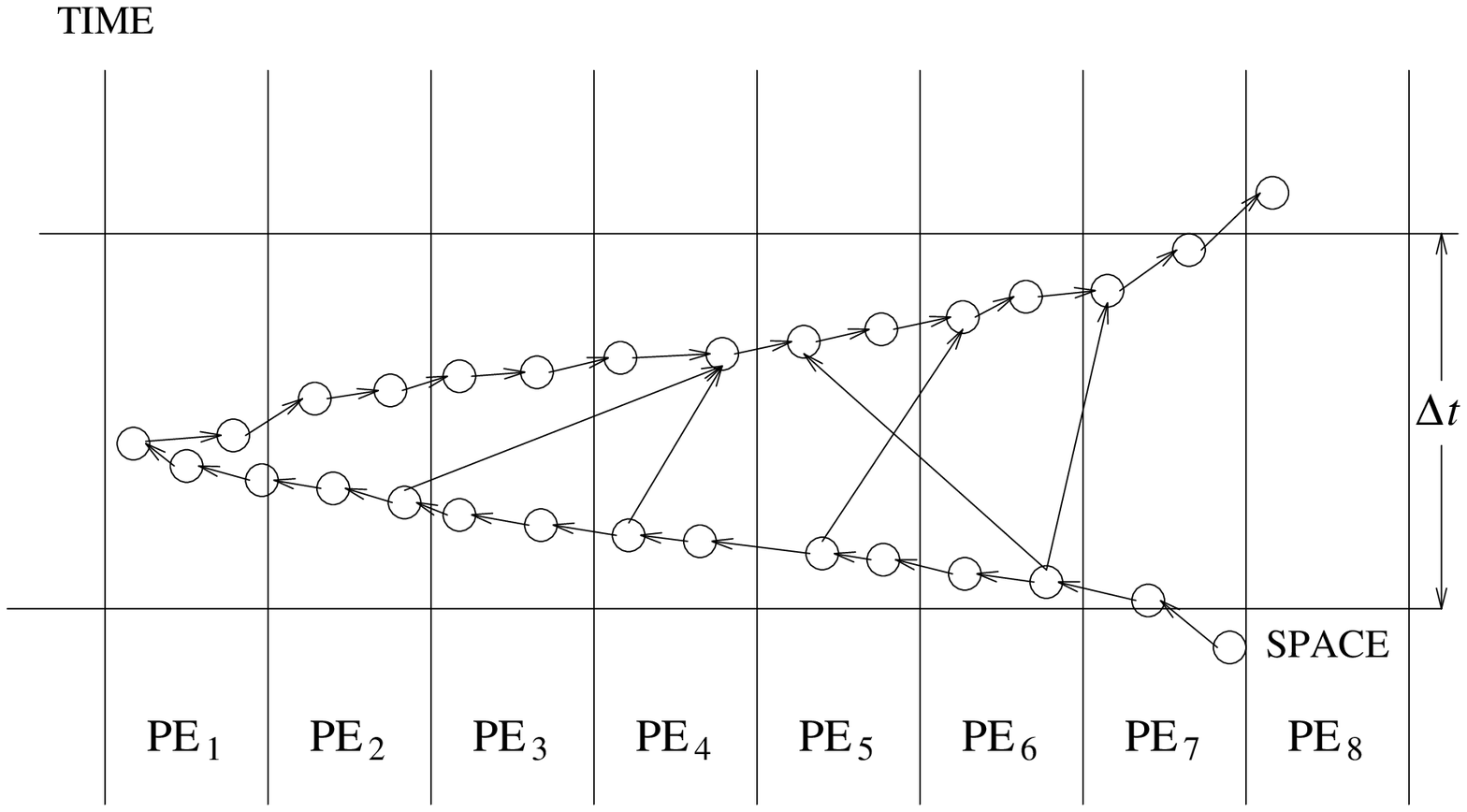,width=7.0in}}
\caption{An event dependency graph which makes inefficient
a parallel execution}
\label{chain}
\end{figure}

\begin{figure}[h]
\centerline{\psfig{file=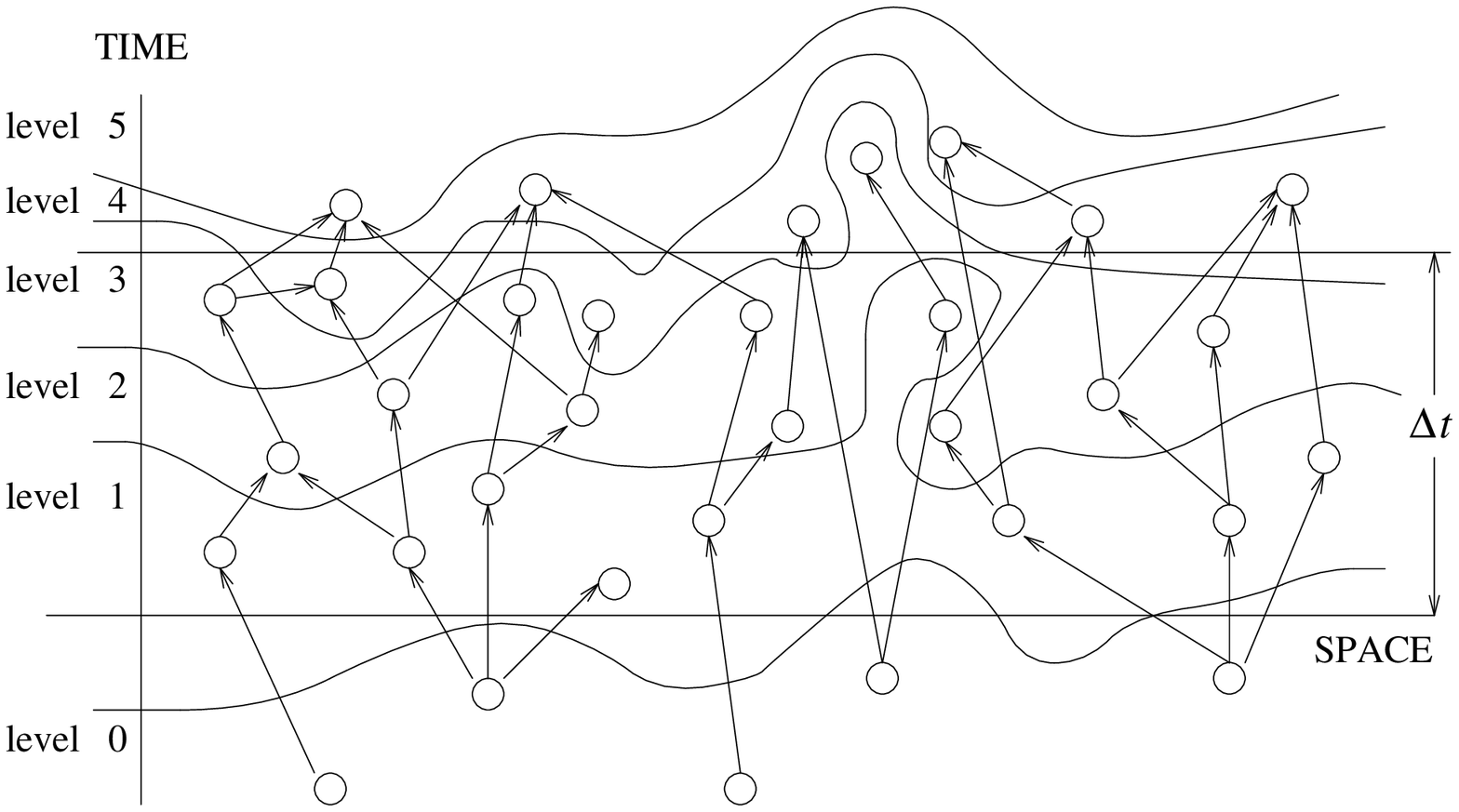,width=7.0in}}
\caption{Event dependency levels. The number of levels
bounds from above the number of iterations of the synchronous
relaxation}
\label{levels}
\end{figure}

Naturally, the efficiency of the synchronous relaxation
method relies
on the convergence to be achieved at each step in
a small number of iterations.
To assess
the number of iterations we examine
the event dependency graph.
For the event chains like the one in Figure~\ref{chain}
there will be many iterations.
\marginpar{\em Figures \ref{chain}, \ref{levels}}
Figure \ref{chain} depicts an artificially difficult, 
worst case example. 
Such event dependency graph may correspond 
to a single indivisible object
which moves in space visiting the areas hosted
by different PEs. 
It is not feasible to make an efficient parallel 
simulation in such a special example.

In Figure \ref{levels}, on the other hand,
an ``average'' example is presented.
It is obtained
by randomly ``sprinkling'' the events-circles and
possible event dependency links
on the time-space diagram,
without a particular application in mind.
A good upper bound on the number of iterations
can be supplied by counting {\em levels}.
The levels can be identified
without knowing the system partitioning into components
hence no such partitioning is shown in Figure~\ref{levels}.
Level 0 
consists of already processed events that are positioned
below the $\Delta t$ strip.
Level 1 consists of those events 
at or above the lower boundary of the strip,
which are immediate effects of only level 0 events.
By induction, level $k$ consists of the events
at or above the lower boundary of the strip,
whose immediate causes are events at levels $k-1, k-2,...1$.
In addition, to qualify for being on level $k$, the event
must have al least one event on level $k-1$ 
among its immediate causes.

Before the step begins,
all level 0 events are correct.
After all the PEs generate their trajectories at iteration 1,
all level 1 events at least will be among the correctly settled events.
It can be seen by induction that
after iteration $k$,
all events on level $k$ or lower are correct.
Thus, the number of levels
for those events of the event dependency graph
that fit within the considered $\Delta t$-strip
is an upper bound on the number of
iterations needed for correctly determining all events
for this strip.

The actual number of iterations can be smaller than this upper
bound for two reasons:
1) initial guesses of events are correct,
2) the event dependency subgraph hosted by a PE
contains a complete set of cause-effects for several
levels without need to know events in the neighboring
PEs.
Situation 1 is not always negligibly rare: in the
applications in which there are only two choices for an event,
reasonable initial guessing might save
iterations.
An extreme case of situation 2 is completely independent subsystems
hosted by different PEs, or, for that matter,
just a single PE which hosts the entire system.
In these conditions,
all events are determined correctly at the first iteration.

The question remains: How many event levels fits
in the $\Delta t$-strip on an ``average''?
A conjecture can be proposed which says,
that, in a ``generic'' example,
if $N$ tends to infinity,
the ``average'' number of levels increases
not faster than $\text{log} N$.
This has been established in several applications,
for example,
in the simulation of circuit-switched networks \cite {EGLW93}.

One may notice a similarity of the synchronous relaxation
algorithm and the Time Warp algorithm \cite{J84}.
Indeed, both algorithms use speculative event processing.
The Time Warp procedure can be qualified as an {\em asynchronous}
relaxation. Instead of frequent synchronization,
the original TW procedure allows each PE to proceed at its own pace,
without explicitly synchronizing with other PEs.
As a result of tighter synchronization,
the synchronous relaxation performs better 
than the TW in a worst case.
The TW is known to sometimes enter 
undesirable modes like rollback avalanche or cascading,
which might slow down unduly even well parallelizable simulations.
Unlike the synchronous relaxation,
no mathematical guarantee of scalability of the TW
algorithm to a large $N$ has been offered.

Whenever there is a choice between a method with speculative
computations and a method without, if
both methods should deliver a scalable parallel simulation,
the non-speculative method should be taken because speculative
computations always involve a heavy computing overhead.
Sometimes for the same simulation model
in some regimes one can do well without speculative
computations, while in the other regimes one can not.
Such is the Ising model example.
The uniformization entails a heavy overhead only for
a low temperature
and then synchronous relaxation is warranted.
For higher temperatures, the 
method discussed in Section \ref{s:sequent}
provides a reasonable alternative.

\section{Conclusion}\label{s:concl}
There are many aspects in computer simulations,
such as visualization, user interface,
convenience and efficiency of programming etc.
The aspect which comes first 
in simulating large dynamic systems 
is that of computing efficiency.
A lesson learned from experiences
in such tasks is that
computing efficiency 
is determined by the properties
of the underlying computational technique
whereas the choice of the best technique 
is not defined by the modeling area.
The same algorithmic idea may work well
across diverse applications and modeling areas.
Another lesson is that no single ``silver bullet'' technique
for efficient simulation has been offered thus far
and that a concrete simulation model may need a combination
of available techniques to work fast.
Sometimes one has to modify the model
to fit a good technique. 
Yet in other cases very substantial improvements in
computing speed are achieved if a basic technique is 
modified or augmented to fit the application,
rather than being used in a fixed ``prepackaged'' form.

\end{document}